\documentstyle[12pt]{article}
\begin{document}
\date{}
\title{Goldstone Bosons\\in the Gaussian Approximation}
\author{Anna Okopi\'nska\\
 Institute of Physics, Warsaw University, Bia\l ystok Branch,\\
 Lipowa 41, 15-424 Bia\l ystok, Poland\\e-mail: rozynek@fuw.edu.pl\\}
\maketitle
\begin{abstract}
\noindent The $O(N)$ symmetric scalar quantum field theory with
$\lambda\Phi^4$ interaction is discussed in the Gaussian approximation.
It is shown that the Goldstone theorem is fulfilled for
arbitrary $N$.
\end{abstract}
\vskip 1 cm
\section{Introduction}
The theory of a real scalar field in n-dimensional Euclidean
space-time with a classical action given by
\begin{equation}
S[\Phi]=\int\![\frac{1}{2}\Phi(x)(-\partial^2+m^2)\Phi(x)+
\lambda (\Phi^{2}(x))^2]\,d^{n}x,
\label{Scl}
\end{equation}
is the most mysterious part of the standard model. Althought
experimentaly not observed, the scalar Higgs field with $m^2 <
0$ and internal $O(4)$ symmetry is necessary to give masses to
interaction bosons in the Weinberg-Salam model of weak
interactions without spoiling renormalizability. Moreover, the
renormalized $\lambda\Phi^4$ theory has been almost rigorously
proved~\cite{tri} to be noninteracting, in contradiction to the
perturbative renormalization, which can be performed order by
order without any signal of triviality. Triviality shows up in
the leading order of the $\frac{1}{N}$ expansion~\cite{BM} in
the $O(N)$ symmetric theory of $N-$component scalar field,
$\Phi(x)=(\Phi_{1}(x),...,\Phi_{N}(x))$. Other non-perturbative
methods, like the Gaussian~\cite{Stev} and
post-Gaussian~\cite{AO,StSt} approximations, have been therefore
applied to study renormalization of the scalar theory in the
case when the number of field components is not large. However,
a serious drawback of the Gaussian approximation for
$N$-component field was an observation~\cite{Con,SAT} that the
Goldstone theorem seemed not to be respected exactly, only in
the limit of $N\rightarrow\infty$ did the would-be Goldstone
bosons become massless. Here we show that this statement is not
true, and is due to a faulty interpretation. In fact, the
Gaussian approximation of the $O(N)$-symmetric theory yields
one massive particle and $(N-1)$ massless Goldstone bosons, in
agreement with the exact result of Goldstone theorem~\cite{Gol}.
There is another claim of existence of Goldstone bosons in the
Gaussian approximation by Dmitrasinovic at al.~\cite{Dmi}, who
found a pole in the four-point Green function of the
$O(2)$-symmetric theory, which was interpreted as a bound state
of two massive elementary excitations. In our work we show that
massless bosons appear in the Gaussian approximation as
elementary fields which are eigenvectors of the one-particle
propagator matrix (two-point Green function).

It is convenient to formulate the approximation method for the
effective action, $\Gamma[\varphi]$, since all the Green's
functions can be obtained in a consistent way, through
differentiation of the approximate expression. A full (inverse)
propagator, required for one-particle states analysis, is given
by the second derivative of the effective action at
$\varphi(x)=\phi_{0}$. The vacuum expectation value of the
scalar field, $\phi_{0}$, can be obtained as a stationary point
of the effective potential, $V(\phi)=-\frac{1}{\int\!
d^{n}x}\left.\Gamma[\varphi]\right|_{\varphi(x)=\phi =const}$.

We shall calculate the effective action, using the optimized
expansion (OE)~\cite{AO}. The method consists in modifying the
classical action of a scalar field~(\ref{Scl}) to the form
\begin{eqnarray}
\lefteqn{S_{\epsilon}[\Phi,G]=\int\!\frac{1}{2}\Phi(x)G^{-1}(x,y)\Phi(y)
\,d^{n}xd^{n}y}\nonumber\\
&&+\epsilon\left[\int\!\frac{1}{2}\Phi(x)[-\partial^2+m^2)\delta(x-y)\!-
\!G^{-1}(x,y)]\Phi(y)\,d^{n}xd^{n}y\!+\!\int\!\lambda (\Phi^{2}(x))^2
\,d^{n}x\right],~~~
\label{Seps}
\end{eqnarray}
with an arbitrary free propagator $G(x,y)$. The effective
action, as a series in an artificial parameter $\epsilon$, can
be obtained as a sum of vacuum one-particle-irreducible diagrams
with Feynman rules of the modified theory. The given order
expression for the effective action is optimized, choosing $G(x,y)$
which fulfills the gap equation
\begin{equation}
\frac{\delta\Gamma_{n}}{\delta G^{-1}(x,y)}= 0,
\label{sta}
\end{equation}
to make the dependence on the unphysical field as weak as
possible.

It has been shown that in the first order of the OE for
one-component field, the inverse of a free propagator can be
taken in the form
\begin{equation}
\Gamma(x,y)=G^{-1}(x,y)=(-\partial^2+\Omega^2(x))\delta(x-y),
\label{pro}
\end{equation}
and the Gaussian effective action (GEA) is obtained~\cite{AO}.
The effective potential, derived from the GEA for a constant
background $\varphi(x)=\phi$, coincides with the Gaussian
effective potential (GEP)~\cite{Stev}, obtained before by
applying the variational method with Gaussian trial functionals
to the functional Schr\"odinger equation.

Here we shall calculate the effective action for $N$-component
field to the first order of the OE. In this case, the inverse of
a trial propagator can be chosen in the form of a symmetric
matrix
\begin{eqnarray}
\Gamma_{i,i}(x,y)&=&(-\partial^2+M_{i}^2(x))\delta(x-y)\nonumber\\
\Gamma_{i,j}(x,y)&=&\Gamma_{j,i}(x,y)=M_{ij}^2(x)\delta(x-y)
\label{prop}
\end{eqnarray}
where the functions $M_{i}^2(x)$ and $M_{ij}^2(x)$ are variational
parameters. The calculation of the effective action can be
simplified, using the observation of Stevenson at al.~\cite{SAT}
that for an $O(N)$ symmetric theory only the shift
$\varphi(x)=(\varphi_{1}(x),...,\varphi_{N}(x))$ of the field
sets a direction in the $O(N)$ space. Thus, the eigendirection
of a free propagator matrix will be radial and transverse, and
the variational parameters for the transverse fields should be
equal, because of the remaining $O(N-1)$ symmetry. In the
coordinate system, in which the shift $\varphi$ points in the
$i=1$ direction, the (inverse) trial propagator can be chosen in
the form of a diagonal matrix with
\begin{eqnarray}
\Gamma_{11}(x,y)&=&G^{-1}(x,y)=(-\partial^2+\Omega^2(x))\delta(x-y)\nonumber\\
\Gamma_{ii}(x,y)&=&g^{-1}(x,y)=(-\partial^2+\omega^2(x))\delta(x-y)
\mbox{~~~~~~for~}i\neq 1,
\end{eqnarray}
and the effective action in the first order of the OE is obtained in the form
\begin{eqnarray}
\Gamma[\varphi]&=&-\int[\frac{1}{2}\varphi(x)(-\partial^2
+m^2)\varphi(x)+\lambda(\varphi^{2}(x))^2]\,d^{n}x
-\frac{1}{2}TrLn G^{-1}\nonumber\\&-&\frac{N-1}{2}TrLn g^{-1} +
\frac{1}{2}\int(\Omega^2(x)-m^2-12\lambda\varphi^{2}(x)) G(x,x)
\,d^{n}x\nonumber\\&+& \frac{(N-1)}{2}\int(\omega^2(x)-m^2-
4\lambda\varphi^{2}(x))
g(x,x)\,d^{n}x-3\lambda\int G^{2}(x,x)\,d^{n}x\nonumber\\&-
&(N^2-1)\lambda\int g^{2}(x,x)\,d^{n}x-2(N-1)\lambda\int G(x,x) g(x,x)\,d^{n}x.
\label{GEA}
\end{eqnarray}
\noindent Requiring the effective action to be stationarity with
respect to small changes of variational parameters
\begin{eqnarray}
\frac{\delta\Gamma}{\delta\Omega^2}=
\frac{\delta\Gamma}{\delta\omega^2}=0,
\end{eqnarray}
results in gap equations
\begin{eqnarray}
\Omega^{2}(x)-m^2-12\lambda{\bf\varphi}^{2}(x)-12\lambda G(x,x)-4(N-1)\lambda
g(x,x)&=&0\nonumber\\
\omega^{2}(x)-m^2-4\lambda{\bf\varphi}^{2}(x)-4\lambda
G(x,x)-4(N+1)\lambda g(x,x)&=&0
\label{gap}
\end{eqnarray}
which determine the functionals $\Omega[\varphi]$ and
$\omega[\varphi]$. When limited to a constant background
$\phi=(\phi_{1},...,\phi_{N})$, the the GEA for $N$-component
field gives the effective potential
\begin{eqnarray}
V(\phi)&=&\frac{m^2}{2}\phi^2+\lambda(\phi^{2})^2+I_{1}(\Omega)+
(N-1)I_{1}(\omega)+\frac{1}{2}(m^2-\Omega^2+12\lambda\phi^{2})
I_{0}(\Omega)\nonumber\\&+&\frac{N-1}{2}(m^2-\omega^2+4\lambda\phi^{2})
I_{0}(\omega)+3\lambda I_{0}(\Omega)^2+(N^2-1)\lambda I_{0}(\omega)^{2}
\nonumber\\&+&2(N-1)\lambda I_{0}(\Omega) I_{0}(\omega)
\label{GEP}
\end{eqnarray}
with the functions $\Omega(\phi)$ and $\omega(\phi)$ determined
by algebraic equations
\begin{eqnarray}
\Omega^{2}-m^2-12\lambda\phi^{2}-12\lambda I_{0}(\Omega)-4\lambda (N-1) I_{0}
(\omega)&=&0,\nonumber\\
\omega^{2}-m^2-4\lambda\phi^{2}-4\lambda I_{0}(\Omega)-4(N+1)\lambda I_{0}
(\omega)&=&0,
\label{gapIp}
\end{eqnarray}
where
\begin{eqnarray}
I_{1}(\Omega)&=&\frac{1}{2}\int\!\frac{d^{n}p}{(2\pi)^{n}}
\ln (p^2+\Omega ^2)\nonumber\\
I_{0}(\Omega)&=&\int\!\frac{d^{n}p}{(2\pi)^{n}}\frac{1}{p^2+\Omega ^2}.
\end{eqnarray}
The same result for the $O(N)$ symmetric GEP was obtained before
in the Schr\"odinger approach~\cite{SAT}. In the OE, a
generalisation of the GEP to space-time dependent fields, the
GEA~(\ref{GEA}), has been obtained. It enables us to derive not
only the effective potential, but also one-particle-irreducible
Green's functions at arbitrary external momenta in the Gaussian
approximation.

The minimum of the GEP is at $\phi_{0}$ fulfilling
\begin{equation}
\frac{\partial V}{\partial\phi_{i}}=(m^2+4\lambda{\bf \phi}^{2}+
12\lambda I_{0}(\Omega)+4(N-1)\lambda I_{0}(\omega)){\bf \phi}_{i}=0;
\label{Vprim}
\end{equation}
therefore, in the unsymmetric minimum we have
\begin{equation}
B=m^2+4\lambda\phi^{2}+12\lambda I_{0}(\Omega)+4(N-1)\lambda I_{0}(\omega)=0.
\label{fimin}
\end{equation}
In the GEP analysis for $N=2$, it was pointed out by Brihaye and
Consoli~\cite{Con} that $\omega[{\bf \phi}_{0}]$ is not equal to
zero, which was interpreted as a violation of Goldstone theorem
in the Gaussian approximation. For the same reason, Stevenson,
All\`es and Tarrach~\cite{SAT} admitted that also for a general
$N$ the Gaussian approximation does not respect the Goldstone
theorem. We would like to point out that this conclusion is
unjustified, for $\Omega$ and $\omega$ are only variational
parameters in the free propagator, and do not correspond to
physical masses of scalar particles. The physical masses have
to be determined as poles of a full propagator in the discussed
approximation. The inverse of that propagator can be obtained as
a second derivative of the GEA~(\ref{GEA}) with an implicit
dependence, $\Omega^2[{bf\varphi}]$ and
$\omega^2[{\bf\varphi}]$, taken into account by differentiation
of the gap equations~(\ref{gap}). Upon performing the Fourier
transform, the two-point vertex is calculated to be
\begin{eqnarray}
\Gamma_{11}(p)&=&\left.\widehat{\frac{\delta^{2}\Gamma}
{\delta\varphi_{1}^2}}\right|_{\varphi(x)=\phi_{0}}=
p^2+m^2+4\lambda\phi^{2}+12\lambda I_{0}(\Omega)+
4(N-1)\lambda I_{0}(\omega)+8\lambda\phi^{2}_{1} A(p)\nonumber\\
\Gamma_{ii}(p)&=&\left.\widehat{\frac{\delta^{2}\Gamma}
{\delta\varphi_{i}^2}}\right|_{\varphi(x)=\phi_{0}}=
p^2+m^2+4\lambda\phi^{2}+12\lambda I_{0}(\Omega)+4(N-1)\lambda I_{0}(\omega)+
8\lambda\phi^{2}_{i} A(p)\nonumber\\
\Gamma_{ij}(p)&=&\Gamma_{ji}(p)=\left.\frac{1}{2}\widehat{\frac{\delta^{2}
\Gamma}{\delta\varphi_{i}\delta\varphi_{j}}}\right|_{\varphi(x)=
\phi_{0}}=8\lambda\phi_{i}\phi_{j} A(p),
\label{propa}
\end{eqnarray}
\noindent where
\begin{eqnarray}
A(p)=1-\frac{18\lambda I_{-1}(\Omega,p)+2\lambda (N-1)
I_{-1}(\omega,p)+24\lambda^2 (N+2) I_{-1}(\Omega,p) I_{-1}(\omega,p)}
{1+6\lambda I_{-1}(\Omega,p)+2\lambda (N+1) I_{-1}(\omega,p)+
32\lambda^2 (N+2) I_{-1}(\Omega,p) I_{-1}(\omega,p)}.
\end{eqnarray}
and
\begin{equation}
I_{-1}(\Omega,p)=2\int\!\frac{d^{n}q}{(2\pi)^{n}}\frac{1}{(q^2+\Omega ^2)
((p+q)^2+\Omega ^2)}.
\end{equation}
Upon diagonalization of the matrix~(\ref{propa}) we obtain an
inverse propagator \mbox{$\gamma_{1}(p)=p^2+B+8\lambda A(p){\bf
\phi}_{0}^2$}, which corresponds to massive particle, and \mbox{$(N-1)$}
inverse propagators \mbox{$\gamma_{i}(p)=p^2+B$} of Goldstone bosons,
since B=0 in the unsymmetric minimum~(\ref{fimin}). Therefore,
for any $N$ the Gaussian approximation of the $O(N)$ symmetric
theory does fully respect Goldstone theorem at the
unrenormalized level.

I would like to acknowledge fruitful discussions with
P. Stevenson and M. Consoli and the warm hospitality of
Instituto Nazionale di Fisica Nucleare in Catania.

\end{document}